\documentclass[preprint,12pt]{elsarticle}

\usepackage{epsfig}
\usepackage{enumerate}

\usepackage{amssymb,amsmath,graphics,graphicx}  
\usepackage{xcolor}

\journal{Elsevier}

\begin{document}

\begin{frontmatter}


\title{When cardinals strategize: An agent-based model of influence and ideology for the papal conclave}



\author{Nuno Crokidakis}
\ead{nunocrokidakis@id.uff.br}

\address{
Instituto de F\'{\i}sica, Universidade Federal Fluminense, Niter\'oi, Rio de Janeiro, Brazil
}

\begin{abstract}
We propose and analyze agent-based models to investigate consensus formation under qualified majority constraints, motivated by the dynamics of papal conclaves. The framework incorporates social imitation, strategic adaptation based on candidate viability and ideological alignment among agents. We consider both a baseline model without explicit ideological structure and an extended version with competing ideological blocs. Numerical simulations show that ideological polarization tends to delay consensus, while increased strategic responsiveness can significantly accelerate convergence. The interplay between social influence and strategic behavior leads to nontrivial collective dynamics, with strong sensitivity of convergence times to model parameters. We further validate the model by comparing its predictions with historical data from papal conclaves held between 1939 and 2025, finding good agreement across a wide range of scenarios. The rapid outcome of the 2025 conclave suggests that informal consensus-building mechanisms may play a key role in accelerating coordination. Beyond this specific application, the proposed framework captures general features of collective decision-making in structured populations and provides a simple platform to study nonlinear consensus dynamics. Our results indicate that small behavioral changes may produce large-scale effects, consistent with phenomena analogous to continuous phase transitions.

\end{abstract}


\end{frontmatter}

%\linenumbers


\section{Introduction}

\quad On May 8, 2025, Cardinal Robert Francis Prevost was elected as Pope Leo XIV after just four rounds of voting, making the 2025 conclave one of the fastest papal elections in nearly two centuries \cite{bbc2}. This swift consensus among the 133 cardinal electors underscores the complex interplay of ideological alignment, strategic behavior and social influence within the College of Cardinals.

The secrecy and ritual surrounding papal conclaves have long attracted public attention - and recently, fictional portrayals as well. The 2024 film Conclave, directed by Edward Berger and released on Prime Video, dramatizes the psychological and political tensions that shape the election of a new pope \cite{movie}. Although fictional, such depictions underscore the complexity and intrigue of real-world conclave dynamics.

From a scientific perspective, the conclave dynamics presents a compelling case for computational modeling. In particular, agent-based models (ABMs) are well suited to explore how individual behavior, social influence and institutional rules interact in collective decision-making processes \cite{abm}. In this context, models of opinion dynamics, including voter-like processes and their constrained variants, have been widely used to investigate consensus formation and polarization in interacting populations \cite{rmp,vazquez2004}. More broadly, recent advances in social physics and complex systems have emphasized the role of agent-based approaches in understanding emergent collective behavior in structured populations \cite{jusup2022,perc2019,bianconi2023}. While ABMs have been widely applied to political, economic, and organizational settings, the conclave offers a unique testbed: a bounded electorate, repeated anonymous voting rounds, a qualified majority requirement (two-thirds) and potential ideological factionalism.

Recent theoretical work has examined the conclave from an axiomatic perspective. In particular, Doghmi et al. \cite{doghmi} analyzed the institutional rules of the papal election through the lens of social choice theory, identifying which axioms, such as anonymity, neutrality and non-dictatorship, are satisfied by the two-thirds supermajority rule and the iterative voting structure of the conclave. Their findings highlight how the conclave balances collective legitimacy with strong majority requirements. While their approach is static and rule-focused, our work offers a complementary perspective by modeling the dynamic process through which consensus may emerge. By simulating agent interactions, strategic adaptations and ideological constraints, we seek to understand not only the fairness of the mechanism, but also its efficiency and responsiveness under varying social conditions.

Recent advancements in computational social science have leveraged agent-based models to simulate and forecast electoral outcomes. Gao et al. \cite{gao} developed an ABM platform that successfully predicted election results by modeling voter interactions, demonstrating the efficacy of such models in capturing complex electoral dynamics. Similarly, Atkinson et al. \cite{atkison} introduced a simple yet flexible ABM to analyze the effects of various voting methods and institutional frameworks on election outcomes. These studies underscore the versatility of ABMs in exploring electoral processes. Building upon this foundation, our research applies ABM techniques to the unique context of the papal conclave, aiming to elucidate the interplay of ideology, influence and strategic behavior in this distinctive electoral setting.

Recent work has also begun to explore computational simulations of papal elections beyond formal or axiomatic frameworks. One notable example is the open-source project ConclaveSim \cite{zhu}, which uses agent-based simulation powered by language models to emulate the 2025 conclave. Each cardinal is represented by a generative agent with ideological priors and communication behavior, allowing for the exploration of negotiation dynamics and alliance formation within the College of Cardinals. While this approach emphasizes narrative modeling and generative reasoning, our work takes a complementary direction by focusing on minimalist, rule-based agent interactions to isolate the effects of social influence, viability perception, and ideological structure on consensus formation. Recent computational studies have begun to model the papal conclave using advanced natural language processing techniques. The authors in \cite{antonioni} developed a framework that analyzes public statements of cardinal electors to map ideological proximities and simulate voting dynamics. Their findings underscore the potential of computational methods in forecasting conclave outcomes and understanding the underlying ideological structures.  Together with axiomatic analyses \cite{doghmi}, these contributions reflect a growing interdisciplinary interest in understanding conclave dynamics through formal and computational lenses.

Building upon such approaches, our study employs agent-based modeling to explore the dynamics of consensus formation within the College of Cardinals. We develop and analyze two models of the papal conclave. The first assumes no explicit ideological structure and focuses on the interplay between social imitation, perceived candidate viability and useful voting. The second extends this framework by assigning cardinals and candidates to ideological blocs (progressives and conservatives), thereby introducing structured preferences and ideological loyalty. By simulating both scenarios, we examine how group composition, strategic adaptation and social mechanisms affect the likelihood and timing of papal election. In addition to exploring the recent 2025 conclave, we calibrate our model using historical data from previous conclaves (1939-2013), adjusting ideological composition and behavioral parameters to reproduce the observed variation in voting rounds. This historical alignment demonstrates the model's explanatory power and highlights how small changes in strategic responsiveness or ideological balance can generate large differences in collective outcomes.

% ##############################################################

\section{Model description}

\subsection{Basic agent-based model}

\quad We consider an agent-based model to describe the dynamics of papal conclave elections. The system consists of $N$ agents, representing cardinals, and $C$ fixed candidates. Each cardinal holds a current voting preference, which may change over time due to social influence and strategic considerations. The election proceeds in discrete rounds, each composed of a voting phase and an opinion update phase.

\begin{enumerate}

\item Voting phase: In each round, every cardinal casts a vote for their current preferred candidate. A candidate is elected if they receive at least two-thirds of the total votes ($\geq (2/3)N$), as required by real conclave rules. If no candidate reaches this threshold, the process continues to the next round.

\item Opinion update phase: After each round, every cardinal may revise their voting preference according to the following rules:

\begin{itemize}

\item Social imitation: with probability $p$, a cardinal adopts the voting preference of a randomly chosen peer.

\item Viability shift: with probability $q$, a cardinal switches support to the most voted candidate in the previous round.

\item Useful voting: if a cardinal's current preferred candidate received less than $5\%$ of the votes, they abandon their choice and adopt the most voted candidate, regardless of prior preference.

\end{itemize}

\end{enumerate}
  
The combination of social imitation and strategic shifts creates a dynamic landscape in which convergence toward a qualified majority may or may not be achieved, depending on the model parameters and initial conditions.

%%%%%%%%%%%%%%%%

\subsection{Model with ideological blocs}

\quad To explore the impact of ideological alignment, we extend the basic model by introducing two ideological groups: progressives and conservatives. Each cardinal is assigned to one of the two groups, and each candidate is also given an ideological label.

\begin{enumerate}

\item Initial conditions: A fraction $f$ of the cardinals is conservative, while the remaining $1-f$ are progressive. In our simulations, we consider a baseline of $f=0.20$, reflecting the estimated real-world composition of the College of Cardinals \cite{bbc}. Candidates are initialized with ideologies drawn independently from a distribution with $20\%$ probability of being conservative.

\item Modified voting behavior: Cardinals preferentially vote for candidates that share their ideology. The initial preference of each cardinal is randomly chosen among the candidates of the same group. However, opinion updates may override this alignment:

\begin{itemize}

\item Imitation and viability updates proceed as before, potentially crossing ideological lines.

\item The useful voting considers the ideological bloc (no shift).
%  Useful voting allows cardinals to abandon ideologically aligned but nonviable candidates in favor of more popular alternatives, including those from opposing blocs.

\end{itemize}

\end{enumerate}

This ideological structure introduces the possibility of prolonged deadlocks and vote clustering within ideological communities. The dynamics balance group loyalty with strategic adaptation, making the convergence to a two-thirds majority more complex. Except for the initial ideological alignment and constrained useful voting, the opinion update rules are identical to those described in Section 2.1.

The simulations begin with a random initial configuration: each of the $N$ cardinals is assigned a preferred candidate selected uniformly at random from the $C$ available options. At each discrete time step (or round), all cardinals cast their vote according to their current preference. If any candidate obtains at least two-thirds of the votes, the election ends and that candidate is declared the winner. Otherwise, a new round begins, during which all cardinals update their preferences based on the rules described earlier - namely, social imitation, response to perceived viability, and, where applicable, useful voting and ideological affinity. This process repeats until either a candidate is elected or the simulation reaches a maximum number of rounds  $T_{max}$. The parameter settings used in the simulations are summarized in Table 1.

%%%%%%%%%%%%%%%%%%%%%%%%%%%%%%%%%%%%%%%%%%%%%%%%%%%%%%%%%%%%%%%%%%%%%%%%
\begin{table}[htbp]
\centering
\resizebox{\textwidth}{!}{  
\begin{tabular}{|c|l|c|}
\hline
Symbol & Description & Typical value
\\
\hline
$N$ & Total number of cardinals & 133  \\ 
$C$ & Number of candidates & 10  \\ 
$p$  & Probability that a cardinal imitates another randomly selected cardinal & 0.1  \\ 
$q$  & Probability that a cardinal switches to the most voted candidate & 0.2  \\ 
$T_{max}$ & Maximum number of rounds allowed before termination & 100   \\
$f$ & Fraction of conservative cardinals (ideological parameter) & 0.20  \\ 
\hline
\end{tabular}
}
\caption{Model parameters used in the simulations of the 2025 conclave, including the number of cardinals $N$, number of candidates $C$, and the parameters controlling social imitation, strategic voting, and ideological composition.}
\label{Tab1}
\end{table}
%%%%%%%%%%%%%%%%%%%%%%%%%%%%%%%%%%%%%%%%%%%%%%%%%%%%%%%%%%%%%%%%%%%%%%%%

%%%%%%%%%%%%%%%%%%%%%%%%%%%%%%%%%%%%%%%%%%%%%%%%%%%%%%%%%%%%%%%%%%%%%%%%

\section{Results}

\subsection{2025 conclave}

\begin{figure}[htp]
\centering
\fbox{\begin{minipage}{36em}
\small
%{\bf Typical output of the numerical code
%} 

\vspace{5mm}

\textbf{Voting round 1:} Candidate 1 = 11 votes;  Candidate 2 = 10 votes;  Candidate 3 = 10 votes;  Candidate 4 = 17 votes;  Candidate 5 = 12 votes;  Candidate 6 = 19 votes;  Candidate 7 = 18 votes;  Candidate 8 = 11 votes;  Candidate 9 = 13 votes;  Candidate 10 = 12 votes  \\

\textbf{Voting round 2:} Candidate 1 = 8 votes;  Candidate 2 = 8 votes;  Candidate 3 = 7 votes;  Candidate 4 = 11 votes;  Candidate 5 = 11 votes;  Candidate 6 = 44 votes;  Candidate 7 = 15 votes;  Candidate 8 = 9 votes;  Candidate 9 = 8 votes;  Candidate 10 = 12 votes   \\

\textbf{Voting round 3:} Candidate 1 = 10 votes;  Candidate 2 = 5 votes;  Candidate 3 = 6 votes;  Candidate 4 = 8 votes;  Candidate 5 = 8 votes;  Candidate 6 = 64 votes;  Candidate 7 = 11 votes;  Candidate 8 = 6 votes;  Candidate 9 = 9 votes;  Candidate 10 = 6 votes  \\

\textbf{Voting round 4:} Candidate 1 = 6 votes;  Candidate 2 = 1 votes;  Candidate 3 = 1 votes;  Candidate 4 = 5 votes;  Candidate 5 = 5 votes;  Candidate 6 = 97 votes;  Candidate 7 = 9 votes;  Candidate 8 = 1 votes;  Candidate 9 = 7 votes;  Candidate 10 = 1 votes  \\

\textbf{Candidate 6 elected in voting round 4 with 97 votes.}

\end{minipage}}
\caption{Typical output of the numerical simulations, illustrating the evolution of the voting process and the convergence to a winning candidate in a single realization of the model.}
\label{fig:output}
\end{figure}

A Typical output of the numerical code is exhibted in Fig. \ref{fig:output}. The program shows the distribuition of votes for each voting round, until a cardinal is elected. For this example, we considered the parameter values shown in Table \ref{Tab1}, for the simplest model defined in section 2.1 (model without ideological groups) \footnote{We tested different values of $C$ (not shown) and verified that the qualitative behavior of the system is robust as long as $C<<N$.}.

\begin{figure}[t]
\begin{center}
\vspace{6mm}
\includegraphics[width=0.6\textwidth,angle=0]{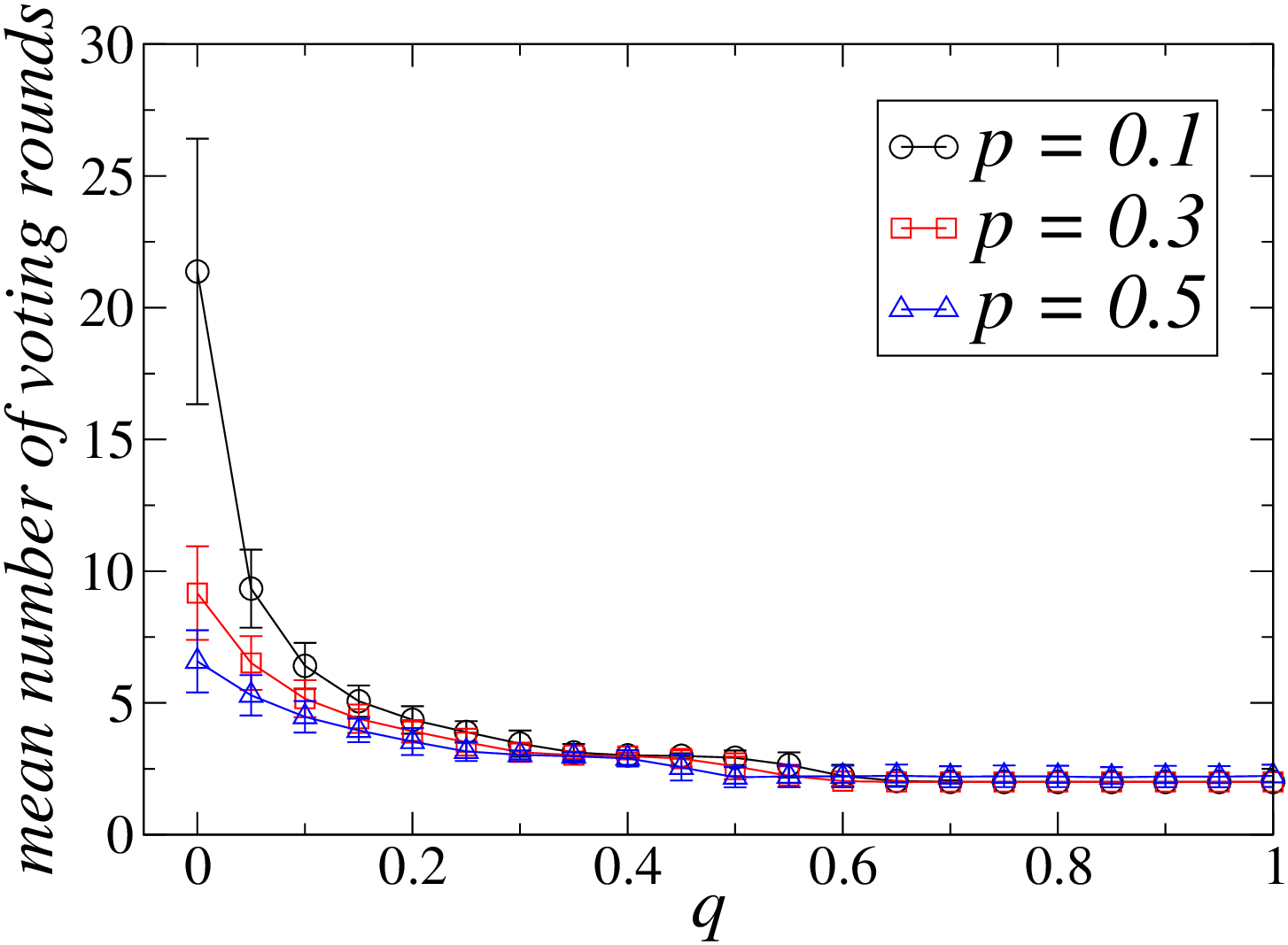}
\end{center}
\caption{Average number of voting rounds until a candidate is elected in the non-ideological model, as a function of the probability $q$ of following the most voted candidate from the previous round. Results are shown for different values of the imitation probability $p$, with fixed $N=133$ cardinals and $C=10$ candidates. Each point represents an average over $1000$ independent simulations. The results show a sharp decrease in the number of rounds with increasing $q$, indicating a coordination transition in which strategic responsiveness significantly enhances convergence efficiency.}
\label{fig2}
\end{figure}

To analyze how social dynamics affect the time required to reach a qualified majority, we simulated the simplified agent-based model without ideological groups, focusing on the interplay between imitation and perceived candidate viability. Figure \ref{fig2} displays the average number of voting rounds until the election of a candidate, as a function of the probability $q$ of adopting the most voted candidate from the previous round. Results are shown for several fixed values of the imitation probability $p$.

For low values of $q$, where cardinals rarely respond to the most viable candidate, the convergence is slow, and the average number of rounds remains high. As $q$ increases, we observe a sharp decline in the number of rounds, indicating a transition from indecisive behavior to coordinated convergence. This reflects a shift from scattered preferences to growing support around a leading candidate, driven by feedback from majority visibility. The presence of even a moderate viability mechanism greatly enhances election efficiency.

Interestingly, higher values of $p$, which correspond to stronger peer imitation, tend to reduce the number of voting rounds required to reach consensus. This suggests that imitation, rather than introducing disruptive noise, facilitates faster convergence by reinforcing dominant local trends and helping clusters of cardinals align more quickly around emerging frontrunners. In this setting, social influence acts as a coordination mechanism, amplifying early asymmetries and promoting the consolidation of votes.

Overall, the model exhibits a realistic coordination transition, in which a small increase in strategic responsiveness $q$ dramatically reduces the expected number of conclave rounds, consistent with known features of collective decision-making in high-consensus environments \cite{rmp}.

%%%%%%%%%%%%%%%%%%%%%%%%%%%%%%%%%%%%%%%%%%%%%%%%%%%%%%%%%%%%%%%%
%%%%%%%%%%%%%%%%%%%%%%%%%%%%%%%%%%%%%%%%%%%%%%%%%%%%%%%%%%%%%%%%

%%%%%%%%%%%%%%%%%%%%%%%%%%%%%%%%%%%%%%%%%%%%%%%%%%%%%%%%%%%%%%%%
\begin{figure}[t]
\begin{center}
\vspace{6mm}
\includegraphics[width=0.6\textwidth,angle=0]{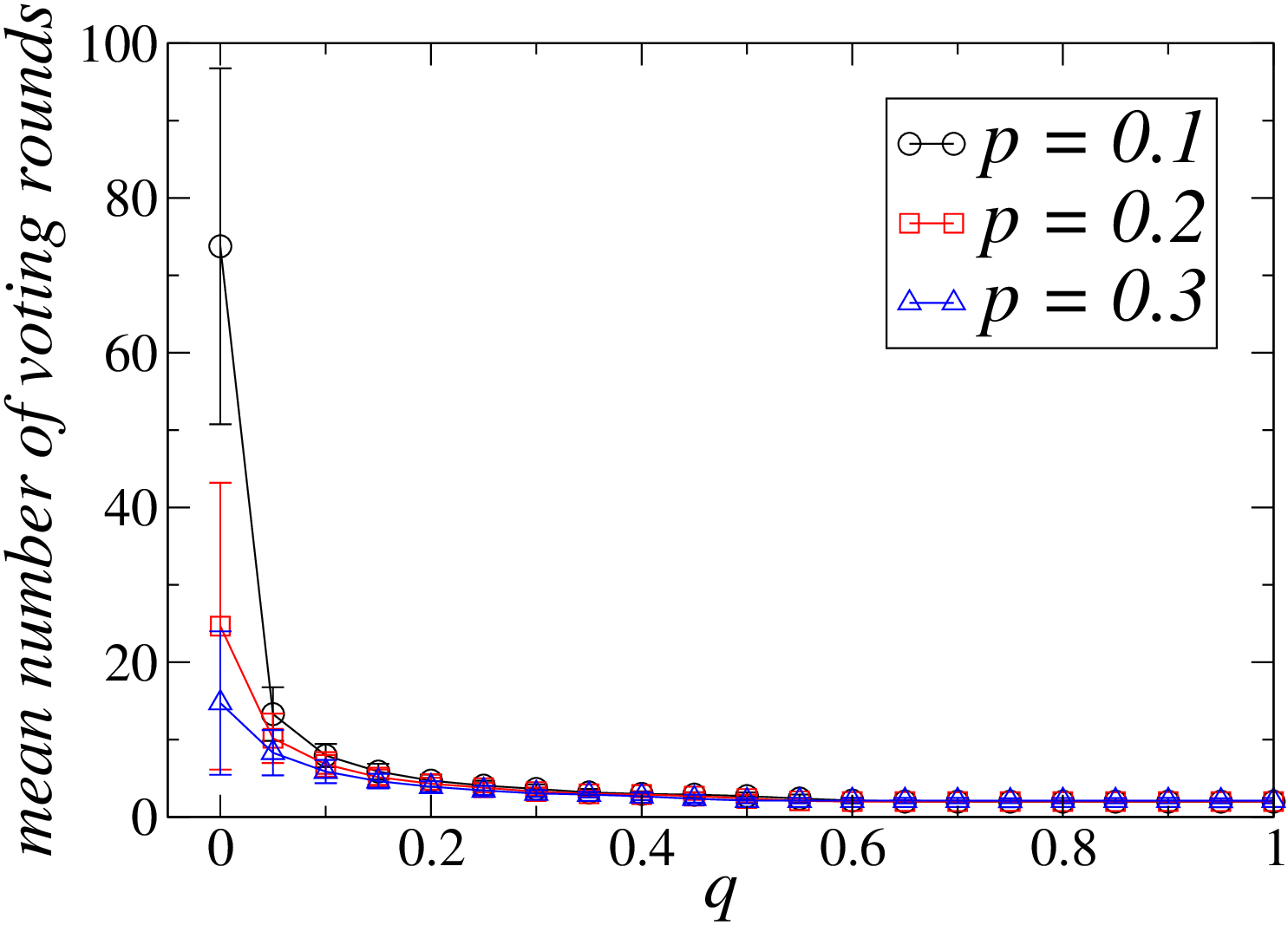}
\end{center}
\caption{Average number of voting rounds in the ideological model, as a function of the probability $q$ of following the most voted candidate. The simulation assumes $N=133$ cardinals and $C=10$ candidates, with $20\%$ of the cardinals initialized as conservatives and $80\%$ as progressives. Results are shown for three values of the imitation probability $p$, and each point represents the average over $1000$ independent simulations. Compared to the non-ideological model, the presence of ideological blocs increases the number of rounds and shifts the coordination threshold to higher values of $q$, reflecting the impact of polarization on consensus formation. Despite this effect, all simulations converge to a successful election within the maximum time $T_{max}$.}
\label{fig3}
\end{figure}
%%%%%%%%%%%%%%%%%%%%%%%%%%%%%%%%%%%%%%%%%%%%%%%%%%%%%%%%%%%%%%%%

%%%%%%%%%%%%%%%%%%%%%%%%%%%%%%%%%%%%%%%%%%%%%%%%%%%%%%%%%%%%%%%%
\begin{figure}[t]
\begin{center}
\vspace{6mm}
\includegraphics[width=0.6\textwidth,angle=0]{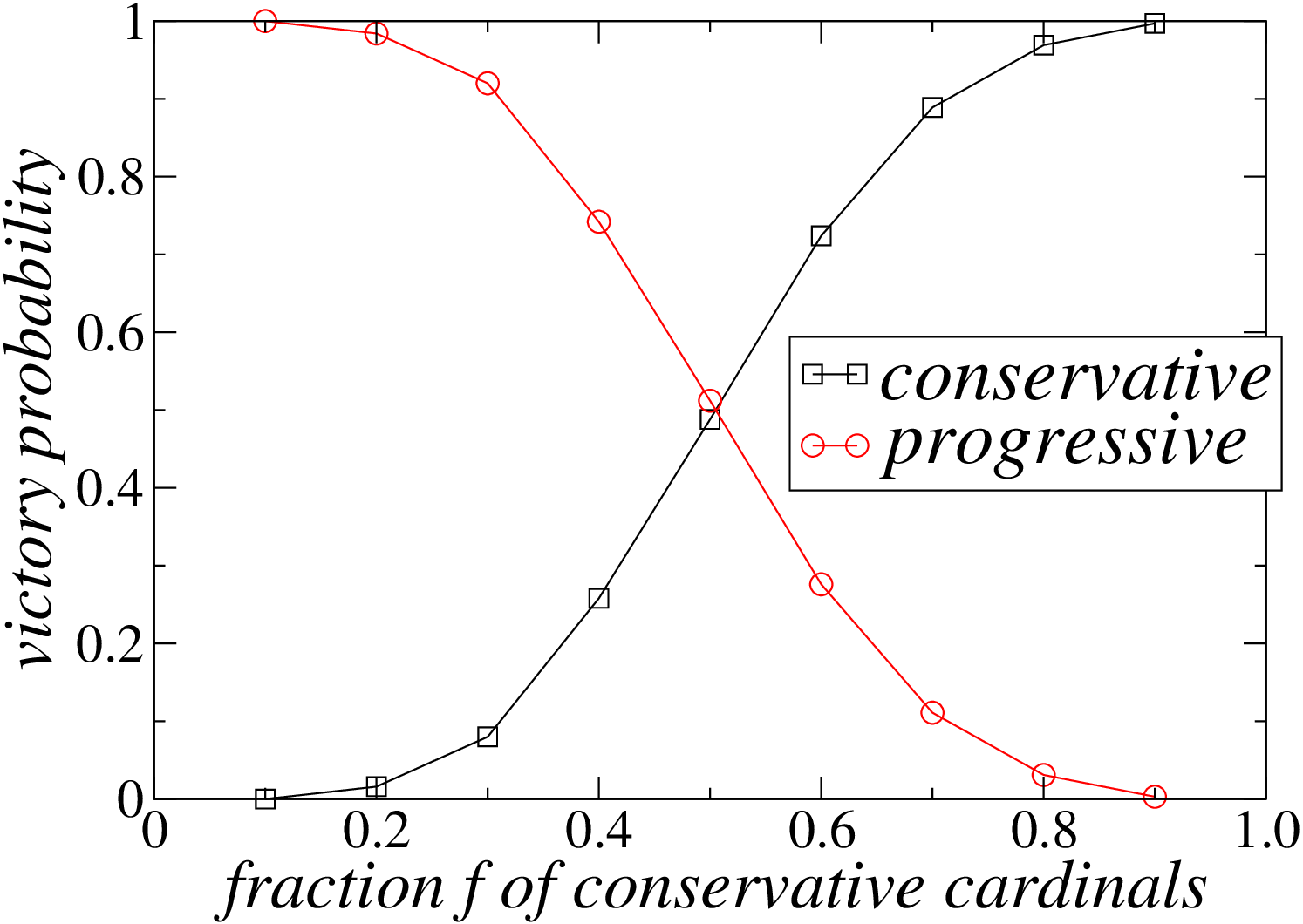}
\end{center}
\caption{Victory probabilities of conservative and progressive candidates as a function of the fraction $f$ of conservative cardinals. Results are based on $1000$ independent simulations per data point, using the ideological model with $N=133$ cardinals and $C = 10$ candidates. Each candidate is randomly assigned an ideological label. The results show a smooth transitions around $f \approx 0.5$, with progressive candidates dominating at low values of $f$, while conservative candidates gain an advantage as their representation increases. This behavior highlights the role of strategic voting and social influence in smoothing the transition between competing ideological blocs.}
\label{fig4}
\end{figure}
%%%%%%%%%%%%%%%%%%%%%%%%%%%%%%%%%%%%%%%%%%%%%%%%%%%%%%%%%%%%%%%%

Figure \ref{fig3} presents the average number of voting rounds until the election of a pope in the ideological version of the model, for a fixed conservative minority fraction of $f=0.20$, a realistic estimate based on recent data on papal appointments. The curves display results for three values of the imitation probability $p$, as a function of the strategic responsiveness parameter $q$.

As expected, the presence of a minority ideological bloc increases the time to consensus relative to the non-ideological model, particularly when $q$ is small. This reflects the reluctance of cardinals to support candidates from opposing ideological groups, which inhibits early convergence. However, as $q$ increases, reflecting stronger responsiveness to candidate viability, the average number of rounds decreases markedly, indicating that even in a polarized setting, strategic behavior can overcome ideological fragmentation.

Notably, higher values of $p$ also contribute to faster convergence in this setting. Unlike in some neutral models where imitation introduces social noise, here it reinforces local consensus within ideological groups, facilitating the eventual coalescence around a viable candidate. These results suggest that, under realistic ideological conditions, a combination of strategic voting and moderate social influence can ensure efficient decision-making, even when a sizable portion of the electorate may initially reject candidates from the dominant bloc.

To investigate how the ideological composition of the conclave influences election outcomes, we ran simulations varying the fraction $f$ of conservative cardinals from 0 to 1. Each point in Figure \ref{fig4} represents the victory probability of conservative and progressive candidates, averaged over $1000$ independent simulations in the ideological model.

The resulting curves reveal a smooth but asymmetric transition between progressive and conservative dominance. When $f<0.4$, progressive candidates win nearly all simulations, as expected from their majority support and ideological alignment. However, conservative candidates only begin to win consistently when $f\geq 0.6$, and the crossover region, where both groups have comparable chances, is relatively narrow. This asymmetry reflects the interplay between ideological loyalty and strategic adaptation: conservative cardinals in the minority face structural disadvantages unless they are able to coordinate effectively or attract cross-ideological support.

This suggests that a moderate conservative minority can still struggle to elect a candidate, even when representing a substantial portion of the electorate. The smoothness of the transition indicates that strategic voting and cross-ideological convergence may be necessary to achieve election under high-consensus constraints. Notably, despite ideological polarization, all simulations resulted in the successful election of a candidate, showing that the model’s combination of social imitation and viability-based adjustment mechanisms is robust enough to avoid deadlock.

% ###############################################################

\subsection{Historical Alignment}

\quad While the primary focus of our simulations was the 2025 conclave, the proposed agent-based model is general and can be retroactively applied to earlier conclaves by adjusting key parameters. Historical records document a wide range of convergence times, from as few as three voting rounds in 1939 (Pius XII) to eleven rounds in 1958 (John XXIII). More recent conclaves, such as those in 2005 and 2013, required four and five rounds, respectively \cite{baumgartner}.

These variations may reflect differences in cardinal composition, ideological polarization, and the extent of informal coordination prior to the formal voting process. For instance, conclaves with a clear ideological majority or a broadly acceptable frontrunner may converge rapidly, while those with more fragmented electorates may take longer. Within our modeling framework, such differences can be captured by adjusting the number of electors $N$, the fraction of conservative cardinals $f$ and the strategic responsiveness parameter $q$.

%%%%%%%%%%%%%%%%%%%%%%%%%%%%%%%%%%%%%%%%%%%%%%%%%%%%%%%%%%%%%%%%%%%%%%%%%%%%
\begin{table}[htbp]
\centering
\resizebox{\textwidth}{!}{  
\begin{tabular}{|c|l|c|p{5.5cm}|}
\hline
\textbf{Conclave Year} & \textbf{Dominant Nominating Pope(s)} & \textbf{Estimated \( f \)} & \textbf{Justification Summary} \\
\hline
1939   & Pius XI  & 0.65 & Majority appointed by Pius XI, traditional orientation. \\
1958   & Pius XII & 0.65 & Most electors appointed by Pius XII. \\
1963   & Pius XII, John XXIII   & 0.60 & Mix of older Pius appointees and some by John XXIII. \\
1978a  & Paul VI & 0.50 & College shaped mostly by Paul VI, moderate reforms. \\
1978b  & Paul VI & 0.40 & After surprise death of John Paul I, limited reshaping. \\
2005   & John Paul II & 0.60 & Clear majority appointed by John Paul II. \\
2013   & John Paul II, Benedict XVI   & 0.30 & Significant Benedict XVI presence, but rising progressives. \\
2025   & Francis  & 0.20 & Majority appointed by Francis with progressive agenda. \\
\hline
\end{tabular}
}
\caption{Estimated fraction of conservative cardinals $f$ in each conclave, based on the appointing history and theological orientation of the dominant pope(s) \cite{pentin}. The values of $f$ are inferred from the composition of the College of Cardinals at the time of each election and are used as input parameters in the simulations presented in this work.}
\label{tab:fractions}
\end{table}
%%%%%%%%%%%%%%%%%%%%%%%%%%%%%%%%%%%%%%%%%%%%%%%%%%%%%%%%%%%%%%%%%%%%%%%%%%%%

%%%%%%%%%%%%%%%%%%%%%%%%%%%%%%%%%%%%%%%%%%%%%%%%%%%%%%%%%%%%%%%%%%%%%%%%
\begin{table*}[tbp]
\begin{center}
\vspace{0.1cm}
\renewcommand\arraystretch{1.2} 
\begin{tabular}{|c|c|c|c|}
\hline
Year & Elected pope & Voting rounds & Cardinals      \\ \hline
2025 & Leo XIV      & 4 & 133  \\ 
2013 & Francis      & 5 & 115  \\ 
2005 & Benedict XVI & 4 & 115 \\
1978b & John Paul II & 8 & 111 \\
1978a & John Paul I  & 4 & 111 \\
1963 & Paul VI      & 6 & 80 \\
1958 & John XXIII   & 11 & 51 \\
1939 & Pius XII     & 3  & 62 \\ \hline
\end{tabular}%
\end{center}
\caption{Data for recent papal conclaves, including the year, elected pope, number of voting rounds, and size of the College of Cardinals at the time of each election \cite{baumgartner,pentin}. These data are used to compare and calibrate the model results with historical observations.}
\label{Tab2}
\end{table*}
%%%%%%%%%%%%%%%%%%%%%%%%%%%%%%%%%%%%%%%%%%%%%%%%%%%%%%%%%%%%%%%%%%%%%%%%

%%%%%%%%%%%%%%%%%%%%%%%%%%%%%%%%%%%%%%%%%%%%%%%%%%%%%%%%%%%%%%%%
\begin{figure}[t]
\begin{center}
\vspace{6mm}
\includegraphics[width=0.6\textwidth,angle=0]{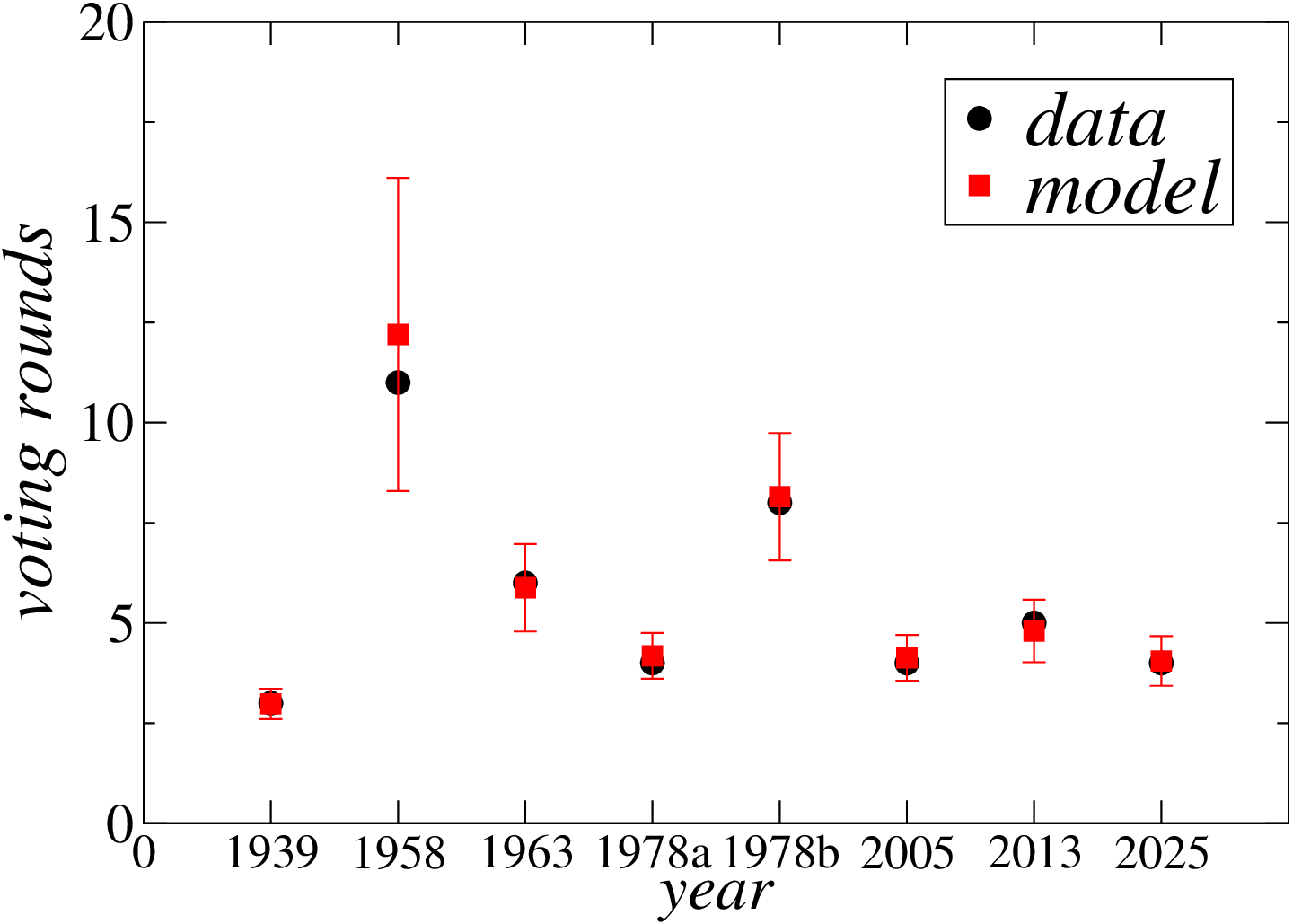}
\end{center}
\caption{Comparison between historical data for papal conclaves, obtained from \cite{baumgartner,pentin}, and results from model simulations. We fixed $p=0.10$ and used inferred values of the conservative fraction $f$, as discussed in the text (see Table \ref{tab:fractions}). The values of the strategic responsiveness parameter $q$ that best fit the historical data are: $q=0.40\, (1939), q=0.05\, (1958), q=0.15\, (1963), q=0.25\, (1978a), q=0.10\, (1978b), q=0.25\, (2005), q=0.20\, (2013)$ and $q=0.25\, (2025)$. Each data point corresponds to an average over $1000$ independent simulation runs. The figure shows that the model is able to reproduce the broad historical variation in conclave durations with good accuracy.}
\label{fig5}
\end{figure}
%%%%%%%%%%%%%%%%%%%%%%%%%%%%%%%%%%%%%%%%%%%%%%%%%%%%%%%%%%%%%%%%

For this comparison, we used the ideological version of the model, adjusting the number of cardinals $N$ and estimating the fraction of conservative cardinals $f$ based on the appointing history of each conclave \cite{baumgartner}. The imitation parameter $p$ was kept fixed at $0.1$, while the viability parameter $q$ was tuned to approximate the observed number of voting rounds. The number of candidates $C$ was kept fixed at 10, consistent with the main simulations. These estimates aim to illustrate the model's flexibility and its ability to reproduce historical variations in convergence time under realistic assumptions.

The estimated fraction of conservative cardinals $f$ used in the simulations was not directly extracted from a single source, but inferred based on the appointing history of each conclave and qualitative assessments available in the literature. Data on the number of electors appointed by each pope, readily available from public ecclesiastical records, was combined with ideological profiles discussed by Pentin \cite{pentin}. Data are exhibited in Table \ref{tab:fractions}.

To further assess the plausibility of our model, we compared its output to historical data from papal conclaves held between 1939 and 2025. Table \ref{Tab2} reports the actual number of voting rounds required to elect a pope in each conclave, alongside key contextual information. Using these data as a benchmark, we performed targeted simulations for each case, keeping the imitation parameter fixed at $p=0.10$ and adjusting the strategic responsiveness $q$ to match the observed convergence times. The values of $f$, representing the estimated fraction of conservative cardinals, were inferred based on the appointing history and ideological orientation of the College of Cardinals, as detailed in Table \ref{tab:fractions}. As shown in Fig. \ref{fig5}, the model is capable of reproducing the observed variation in conclave durations across decades, from the very rapid outcome in 1939 to the more prolonged decision processes in 1958 and 1978b. These results suggest that relatively small changes in strategic responsiveness, interpreted as willingness to abandon ideological rigidity in favor of electability, can account for significant variation in electoral efficiency over time.

% ###############################################################

\section{Final remarks}

\quad The real-world 2025 papal conclave concluded after only four rounds, placing it among the fastest conclaves in modern history. When compared to our simulation results, this outcome aligns with scenarios characterized by high strategic responsiveness and relatively low ideological resistance. In the non-ideological version of the model, convergence within 4 to 6 rounds typically occurs when cardinals strongly prioritize viability (i.e., high $q$) while the imitation probability $p$ remains moderate.

In contrast, the presence of ideological blocs in the extended model generally increases the number of rounds required for consensus, especially under moderate $q$, due to initial polarization and reduced cross-group coordination. The rapid 2025 outcome therefore suggests that ideological divisions were either not strongly expressed during voting or that strategic alignment occurred early, possibly prior to the formal voting process. This interpretation reinforces the idea that informal consensus-building mechanisms, such as pre-conclave discussions \cite{galam2002}, may play a decisive role in accelerating convergence.

Beyond the 2025 case, we extended our analysis by comparing model simulations to historical data from conclaves held between 1939 and 2013. Using estimated ideological fractions $f$ and adjusting $q$ to match observed durations, we found that the model is capable of reproducing the empirical variation in the number of voting rounds with good agreement. These findings support the model's interpretive power and highlight how modest shifts in strategic behavior or ideological composition can yield large-scale differences in election dynamics.

Although the present study is motivated by the specific context of papal conclaves, the proposed framework captures general features of collective decision-making under qualified majority constraints. In particular, the interplay between social influence, strategic adaptation and structured preferences may be relevant to a wide range of systems in which agents must coordinate under strong consensus requirements. In this sense, the model provides a simple but flexible platform for exploring nonlinear dynamics of consensus formation in complex social systems.

Finally, the observed transitions in consensus time and victory probability suggest that conclave dynamics may exhibit features analogous to continuous phase transitions. In particular, small parameter changes, especially in strategic responsiveness, can lead to sharp macroscopic shifts in collective outcomes. This resemblance to phenomena in statistical physics further illustrates the potential of agent-based modeling for capturing the emergent complexity of institutional decision-making \cite{rmp}.

Overall, the proposed framework offers insight into collective decision-making under high-consensus constraints, and may be extended to other institutional settings characterized by ideological factions and strategic voting.

% ###############################################################

\section*{Acknowledgments}
The author acknowledges partial financial support from the Brazilian scientific funding agency Conselho Nacional de Desenvolvimento Cient\'ifico e Tecnol\'ogico (CNPq, Grants 308643/2023-2 and 406820/2025-2).

%% New version of the num-names style
\bibliographystyle{elsarticle-num-names}
%\bibliography{alcoholism.bib}



\end{document}